\documentclass{article}
\usepackage{graphics}
\usepackage{graphicx}
\usepackage{url}
\usepackage{amsfonts}
\usepackage{amsmath}
\usepackage{amssymb}
\usepackage{bbold}
\usepackage{a4wide}
\usepackage{enumitem}
\usepackage{scalerel}

\newcommand{\be}{\begin{equation}}
\newcommand{\bea}{\begin{eqnarray}}
\newcommand{\ee}{\end{equation}}
\newcommand{\eea}{\end{eqnarray}}
\newcommand{\nn}{\nonumber}

\newcommand{\qa}{\alpha}
\newcommand{\qb}{\beta}
\newcommand{\qg}{\gamma}
\newcommand{\qG}{\Gamma}
\newcommand{\qd}{\delta}

\newcommand{\qe}{\varepsilon}

\newcommand{\qy}{\theta}

\newcommand{\qk}{\kappa}
\newcommand{\ql}{\lambda}

\newcommand{\qr}{\rho}
\newcommand{\qs}{\sigma}

\newcommand{\qt}{\tau}
\newcommand{\qf}{\varphi}

\newcommand{\qj}{\psi}
\newcommand{\qJ}{\Psi}
\newcommand{\qo}{\omega}
\newcommand{\qO}{\Omega}

\newcommand{\tr}{{\rm tr}\,}

\newcommand{\dagg}{^{\dag}}

\newcommand{\fr}[2]{{\textstyle \frac{#1}{#2}}}

\newcommand{\EE}{{\mathbb E}}

\newcommand{\one}{{\mathbb 1}}

\newcommand{\bits}{ \{0,1\} }

\newcommand{\cB}{{\mathcal B}}

\newcommand{\cE}{{\mathcal E}}
\newcommand{\cF}{{\mathcal F}}

\newcommand{\cH}{{\mathcal H}}
\newcommand{\cI}{{\mathcal I}}

\newcommand{\cK}{{\mathcal K}}

\newcommand{\cM}{{\mathcal M}}

\newcommand{\cO}{{\mathcal O}}
\newcommand{\cP}{{\mathcal P}}
\newcommand{\cQ}{{\mathcal Q}}

\newcommand{\cS}{{\mathcal S}}
\newcommand{\cT}{{\mathcal T}}
\newcommand{\cU}{{\mathcal U}}
\newcommand{\cV}{{\mathcal V}}

\newcommand{\cX}{{\mathcal X}}

\newcommand{\pr}{{\rm Pr}}

\newcommand{\isdef}{\stackrel{\rm def}{=}}

\newcommand{\ket}[1]{| #1 \rangle}
\newcommand{\bra}[1]{\langle #1 |}

\begin{document}

\newtheorem{theorem}{Theorem}[section]
\newtheorem{property}[theorem]{Property}
\newtheorem{lemma}[theorem]{Lemma}
\newtheorem{definition}[theorem]{Definition}
\newtheorem{corollary}[theorem]{Corollary}
\newtheorem{conjecture}[theorem]{Conjecture}

\setlength{\parindent}{0mm}

\title{Quantum Alice and Silent Bob\\
{\large Qubit-based Quantum Key Recycling with almost no classical communication}}

\author{Daan Leermakers and Boris \v{S}kori\'{c} \\
{\tt\footnotesize d.leermakers.1@tue.nl, b.skoric@tue.nl} }

\date{ }

\maketitle

\begin{abstract}
\noindent 
We answer an open question about Quantum Key Recycling (QKR):
Is it possible to put the message entirely in the qubits without increasing the number of qubits?
We show that this is indeed possible.
We introduce a prepare-and-measure QKR protocol where the communication from Alice to Bob consists
entirely of qubits.
As usual, Bob responds with an authenticated one-bit accept/reject classical message.\footnote{
The title refers to the movie character Silent Bob, who hardly ever speaks.
}

\noindent
Compared to Quantum Key Distribution (QKD), QKR has reduced round complexity.
Compared to previous qubit-wise QKR protocols, our scheme has far less classical communication.

\noindent
We provide a security proof in the universal composability framework
and find that the communication rate is asymptotically 
the same as for QKD with one-way postprocessing. 

\end{abstract}

\section{Introduction}
\label{sec:intro}

\subsection{Quantum Key Recycling}

QKR achieves information-theoretically 
secure communication in such a way that no key material is used up
as long as the quantum channel is undisturbed. 
Compared to QKD followed by classical one-time-pad message encryption, QKR's main advantage is 
{\em reduced round complexity}:
QKR needs only one message from Alice to Bob, and one authenticated bit from Bob to Alice.
QKD needs at least two messages from Alice to Bob.
Furthermore, a minor advantage is that QKR does not discard any qubits, whereas QKD does.

A prepare-and-measure QKR scheme based on qubits was proposed already in 1982~\cite{BBB82}.
Then QKR received little attention for a long time. 
A security proof\footnote{
For a scheme slightly different from  \cite{BBB82}.
} 
for qubit-based\footnote{
as opposed to schemes that work with higher-dimensional spaces, e.g. 
using mutually unbiased bases \cite{DPS2005,DBPS2014}.
} 
QKR was given only in 2017 by Fehr and Salvail~\cite{FehrSalvail2017}.
In \cite{QKR_noise} it was shown (for a scheme similar to \cite{FehrSalvail2017})
that the communication rate in case of a noisy quantum channel
is asymptotically the same as for QKD with one-way postprocessing.

\subsection{Related work; putting the message in the quantum states}
\label{sec:embedding}

Different from the classical setting, in the quantum cryptographic setting authentication implies encryption \cite{Barnum2002}. 
Portmann \cite{Portmann2017} showed that quantum authentication is possible with re-use of all the encryption keys,
but states as an open problem to find a {\em prepare-and-measure} QKR scheme 
for classical messages.

All currently existing qubit-wise prepare-and-measure QKR schemes encode random bits rather than the message into the quantum state, and then extract a classical One-Time Pad (OTP) from these random bits.
Alice sends a classical ciphertext (the message xor'ed with the OTP) along with the quantum states.

In 2003 Gottesman \cite{uncl} proposed a scheme called `Unclonable Encryption'
which encodes a message directly into qubit states. 
Although some of the keys in his schemes can be re-used, still $n$ key bits are discarded when sending an $n$-bit message.
The high-dimensional QKR of Damg{\aa}rd, Pedersen and Salvail \cite{DPS2005,DBPS2014}
has full recycling of keys, but requires quantum computation for encryption and decryption.


\subsection{Contributions}
\label{sec:contrib}

We answer the open question whether it is possible to have prepare-and-measure qubit-based QKR
with the message entirely contained in the qubits, without increasing the number of qubits.
The answer is affirmative.
We present such a scheme; compared to \cite{QKR_noise} Alice's classical message is
removed.

While the contribution of this paper
may not have a great practical impact
(after all the classical channel is a cheap resource compared to the quantum channel),
we find that reducing the overall communication to the bare minimum is of theoretical interest.

\begin{itemize}[leftmargin=4mm,itemsep=0mm]
\item 
Our protocol is a modification of the scheme by Leermakers and \v{S}kori\'{c} \cite{QKR_noise}. 
The main difference lies in the masking of the message and in the privacy amplification.

\item 
In case of {\tt Reject}, Alice and Bob have to tap into fresh key material. 
We implement this key update by hashing fresh key material into the old keys. 
This reduces the {\tt Reject}-case key expenditure
with respect to \cite{FehrSalvail2017} and \cite{QKR_noise}. 
In the absence of noise the {\tt Reject}-case key expenditure asymptotically equals the length of the message, which is optimal \cite{DPS2005}.

\item 
We prove the security of our protocol against general attacks. 
We use a universally composable measure of security, namely
the diamond norm between the actual protocol and an idealized protocol in which 
the secrets are replaced by random strings after protocol execution.
The proof follows the same steps as \cite{QKR_noise}, 
and at an early stage the {\tt Accept}-case part of the proof reduces exactly to the derivation in~\cite{QKR_noise}. Notably, this proof technique achieves optimal rates for 6-state as well as 4-state encodings.

\item
The asymptotic communication rate of our scheme (the number of message bits divided by the number of qubits)
equals 
that of QKD with one-way postprocessing.
The finite-size effects are the same as \cite{QKR_noise}, but with an additional 
small term due to the new key refresh procedure in the {\tt Reject} case. 

\end{itemize}

\subsection{Outline}

In Section~\ref{sec:prelim} we introduce notation and briefly review
post-selection and the results of \cite{QKR_noise}.
We state our motivation in Section~\ref{sec:motivation}, and we
list the steps of the proposed protocol in Section~\ref{sec:protocol}.
Section~\ref{sec:equivalent} presents a stepwise re-formulation of the protocol
which is equivalent in terms of security but better suited to the proof technique.
In Section~\ref{sec:CPTP} we derive the output state of the protocol,
and in Section~\ref{sec:proof} we give the security proof.
We conclude with a discussion and suggestions for future work.

\section{Preliminaries}
\label{sec:prelim}

\subsection{Notation and terminology}
\label{sec:notation}

Classical Random Variables (RVs) are denoted with capital letters, and their realisations
with lowercase letters. The probability that a RV $X$ takes value
$x$ is written as $\pr[X=x]$.
The expectation with respect to RV $X$ is denoted as 
$\EE_x f(x)=\sum_{x\in\cX}\pr[X=x]f(x)$.
Sets are denoted in calligraphic font. 
The notation `$\log$' stands for the logarithm with base~2.
The notation $h$ stands for the binary entropy function $h(p)=p\log\fr1p+(1-p)\log\fr1{1-p}$.
Sometimes we write $h(\{p_1,\ldots,p_k\})$ meaning $\sum_i p_i\log\fr1{p_i}$.
Bitwise XOR of binary strings is written as `$\oplus$'.
The Kronecker delta is denoted as $\qd_{ab}$.
The complement of a bit $b\in\bits$ is written as $\bar b=1-b$. 
The Hamming weight of a binary string $x$ is written as~$|x|$.
We will speak about
`the bit error rate $\qg$ of a quantum channel'.
This is defined as the probability that a classical bit $g$, sent by Alice embedded in a qubit,
arrives at Bob's side as $\bar g$.
We write $\one$ for the identity matrix. 

For quantum states we use Dirac notation.
A qubit state with classical bit $x$ encoded in basis $b$ is written as $\ket{\qj^b_x}$.
We call $x$ the {\em payload}.
We will always assume that we are working with 6-state encoding (known from 6-state QKD, with three possible bases)
or 8-state encoding \cite{SdV2017,LS2018}. 
Occasionally we will comment if a result is different for BB84-encoding.
 
The notation `tr' stands for trace.
Let $A$ have eigenvalues~$\ql_i$. 
The 1-norm of $A$ is written as $\|A\|_1=\tr\sqrt{A\dagg A}=\sum_i|\ql_i|$. The trace distance between matrices $\qr$ and $\qs$ is denoted as $\qd(\qr;\qs)=\frac{1}{2} ||\qr-\qs||_1$. It is a generalisation of the statistical distance and represents the maximum possible advantage one can have in distinguishing $\rho$ from $\qs$. 

Quantum states with non-italic label `A', `B' and `E' indicate the subsystem of Alice/Bob/Eve.
Consider uniform classical variables $X,Y$ and a quantum system under Eve's control that depends on $X$ and~$Y$. 
The combined classical-quantum state is $\qr^{XY \rm E}=\EE_{xy} \ket{xy}\bra{xy} \otimes \qr^{\rm E}_{xy}$. 
The state of a sub-system is obtained by tracing out all the other subspaces, 
e.g. $\qr^{ Y \rm E}={\rm tr}_X \qr^{XY\rm E}=\EE_y \ket y\bra y\otimes\qr^{\rm E}_y$, with $\qr^{\rm E}_y=\EE_x\qr^{\rm E}_{xy}$.
The fully mixed state on Hilbert space $\cH_{\rm A}$ is denoted as~$\chi^{\rm A}$.
The security of the variable $X$, given that Eve holds the `E' subsystem,
can be expressed in terms of a trace distance as follows  \cite{RK2005},
\be
	d(X|{\rm E})\isdef\qd\Big(\qr^{X\rm E} \; ; \;\;
	\chi^X\otimes\qr^{\rm E}\Big)
\ee
i.e.\,the distance between the true classical-quantum state and a state in which 
$X$ is completely unknown to Eve.

We write $\cS(\cH_{\rm A})$ to denote the space of density matrices on the Hilbert space $\cH_{\rm A}$.
 Any quantum channel can be described by a completely positive trace-preserving (CPTP) map 
$\cE: {\cS({\cH_{\rm A}})} \rightarrow {\cS(\cH_{\rm B})}$ that transforms a mixed state $\rho^{\rm A}$ 
to $\rho^{\rm B}$: $\cE(\rho^{\rm A}) = \rho^{\rm B}$.
For a map $\cE: S(\cH_{\rm A}) \rightarrow S(\cH_{\rm B})$, the notation $\cE(\rho^{\rm AC})$ stands for 
$(\cE \otimes \one_C) (\rho^{\rm AC})$, 
i.e.~$\cE$ acts only on the ‘A’ subsystem. 

The diamond norm of $\cE$ is defined as $\| \cE \|_\diamond = \frac12 \sup_{\rho^{\rm AC} \in \cS( \cH_{\rm AC})} \| \cE(\rho^{\rm AC})\|_1$ with ${\cH_{\rm C}}$ an auxiliary system that can be considered to be of the same dimension as $\cH_{\rm A}$. 
The diamond norm $\|\cE-\cE'\|_\diamond$ can be used to upper
bound the probability of distinguishing two CPTP maps $\cE$ and $\cE'$ given that the process is observed once. 
The maximum probability of a correct guess is $\frac12 + \frac14 \| \cE - \cE' \|_\diamond$. 
The security of a protocol is often quantified by the diamond norm between the real protocol 
$\cE$ and an protocol with ideal functionality $\cF$. 
When $\| \cE - \cF \|_\diamond \leq \qe$ 
we can consider $\cE$ to behave ideally except with probability $\qe$; 
this security metric is composable with other (sub-)protocols \cite{TL2017}. 

A family of hash functions $H=\{h:\cX\to\cT  \}$ is called pairwise independent 
(a.k.a.\,2--independent or strongly universal)
\cite{WegmanCarter1981} 
if for all distinct pairs $x,x'\in\cX$
and all pairs $y,y'\in\cT$ it holds that
$\pr_{h\in H}[h(x)=y \wedge  h(x')=y']=|\cT|^{-2}$.
Here the probability is over random~$h\in H$.
Pairwise independence can be achieved  with a hash family of size $|H|=|\cX|$.

\subsection{Definition of QKR rate and security}
\label{sec:defsecurity}

We define the rate of a quantum communication protocol as the number of useful message bits communicated per sent qubit.

Informally, we define QKR security as follows.
Let $k$ denote all the shared keys of Alice and Bob in the current instantiation of the protocol.
Let $\tilde k$  be the keys in the next instantiation, computed in a way that depends on Bob's feedback.
We define the Key Recycling property as follows,
\begin{itemize}[leftmargin=4mm,itemsep=0mm]
\item 
If Bob's feedback message is {\tt Accept}, then $\tilde k$ {\em is computed without tapping into 
new key material}.\footnote{
Note that we do not require re-use of the exact same keys in unmodified form.
}
\end{itemize}
Two security properties must be satisfied:
\begin{itemize}[leftmargin=4mm,itemsep=0mm]
\item 
Even if Eve intercepts the whole ciphertext, she cannot obtain any information about the message.
\item
Using $\tilde k$ in the next instantiation does not endanger any message.
\end{itemize}
Formally, we work with an EPR-state based version of the protocol.
We consider a quantum-classical state $\qr^{K\tilde KM\qO {\rm E}}$
containing the classical random variables $K,\tilde K,M,\qO$, where $M$ is the message
and $\qO$ is Bob's feedback bit,
as well as Eve's quantum side information (the subsystem denoted as `E').
This state is the result of the protocol $\cE$ acting on an input state $\qs$ which represents 
the noisy EPR pairs prepared by Eve, $\qr^{K\tilde KM\qO {\rm E}}=\cE(\qs)$.
An `ideal' version of the protocol is denoted as $\cF$, and we write $\qf^{K\tilde KM\qO {\rm E}}=\cF(\qs)$. 
It satisfies $\qf^{M\qO\rm E}=\qf^M\otimes \qf^{\qO\rm E}$
and $\qf^{\tilde K M\qO \rm E}=\qf^{\tilde K}\otimes\qf^{M\qO\rm E}$.
The first equation expresses the fact that the message $M$ is entirely decoupled from the subsystems available to Eve.
The second says that, even in the case of known plaintext, Eve has no information about the updated key~$\tilde K$.
We say that the QKR scheme $\cE$ is $\qe$-secure if $\| \cE-\cF \|_\diamond\leq\qe$.

\subsection{Post-selection}
\label{sec:post-selection}

For protocols that are invariant under permutation of their inputs it has been shown \cite{CKR2009} that security 
against collective attacks (the same attack applied to each qubit individually)
implies security against general attacks, at the cost of extra privacy amplification. 
Let $\cE$ be a protocol that acts on $S(\cH_{\rm AB}^{\otimes n})$
and let $\cF$ describe the perfect functionality of that protocol.
If for all permutations $\pi$ on the input there exists a map $\cK_\pi$ on the output such that $\cE \circ \pi = \cK_\pi \circ \cE$ then,
\bea
\| \cE -\cF \|_\diamond &\leq& (n+1)^{d^2-1} \max_{\qs \in S(\cH_{\rm ABE})} \Big\| (\cE - \cF)  ( \qs^{\otimes n})\Big\|_1  
\label{eq:post-selection}
\eea
where $d$ is the dimension of the $\cH_{\rm AB}$ space. ($d=4$ for qubits).
The product form $\qs^{\otimes n}$ greatly simplifies the security analysis:
now it suffices to prove security against `collective' attacks,
and to pay a price $2(d^2-1)\log(n+1)$ in the amount of privacy amplification,
i.e.~the output size of the privacy amplification step is reduced by this amount. 

\subsection{Brief summary of results from \cite{QKR_noise}}
\label{sec:prev_QKR}
It was shown that the asymptotic communication rate of QKR is the same as the rate of QKD with one-way postprocessing. 
Alice encodes random bits into the qubits; 
over a classical channel she sends a ciphertext, OTP-encrypted information for error-correction, and an authentication tag. 
Let the CPTP map $\cE$ be the protocol of \cite{QKR_noise},
and $\cF$ its idealized version where the message and the next round's keys are completely unknown to Eve.
It was shown that
\be
	\| \cE -\cF\|_\diamond \leq  
	2^{-\ql+1} 
	+(n+1)^{15}\; \min \big(\qe+\frac12 {\rm tr}_E \sqrt{|\cB|^n2^\ell {\rm tr}_{BS} \big(\bar \rho^{BS {\rm E}}\big)^2}, P_{\rm corr}\big),
\label{DiamondIntermediate}
\ee
where $\ql$ is the length of the authentication tags,
$\qe$ the amount of state `smoothing' \cite{RK2005},
$n$ the number of qubits,
$\cB$ the alphabet of the qubit basis choice,
$\ell$ the message length,
$B$ the basis sequence,
$S$ the random data encoded in the qubits,
and $P_{\rm corr}$ the noise-dependent probability of successful error correction.
The $\bar \qr^{BS\rm E}$ is the state $\cE(\qs^{\otimes n})$ (see Section~\ref{sec:post-selection}) 
smoothened by an amount~$\qe$,
with everything traced out except the $B$, $S$ and E subsystems.
If 6-state encoding\footnote{
For 4-state (BB84) `conjugate' coding Eve has two degrees of freedom, i.e. a more powerful attack.
} 
of bits is used then
the $4\times4$ matrix $\qs$ is completely determined \cite{RennerThesis} by a single parameter: 
the bit error probability~$\qg$ on the quantum channel. 
Asymptotically for large $n$, the bound (\ref{DiamondIntermediate}) reduces to  
\be
	\| \cE -\cF\|_\diamond \leq  2^{-\ql+1} +
	n^{15}\;\min \big(\sqrt{2^{\ell-n+n h(\{1-\frac32 \qg,\frac\qg2 ,\frac\qg2,\frac\qg2\})
	-nh(\qg)}},P_{\rm corr}\big),
\label{OldAsymptotic}
\ee
which yields exactly the same rate\footnote{
The term $nh(\qg)$ gets cancelled because Alice and Bob expend $nh(\qg)$ bits of key material
to OTP the redundancy bits.
} 
$1-h(\{1-\frac32 \qg,\frac\qg2 ,\frac\qg2,\frac\qg2\})$ as 6-state QKD with one-way postprocessing.\footnote{
For 4-state encoding the result is different from (\ref{OldAsymptotic}) and yields the BB84 rate.
}
The security of $N$ QKR rounds follows from
$\| \cE_N\circ\cdots\circ\cE_1-\cF_N\circ\cdots\circ\cF_1 \|_\diamond  \leq N\| \cE-\cF \|_\diamond$.

\section{Motivation}
\label{sec:motivation}

As mentioned in Section~\ref{sec:intro}, current QKR schemes all have some drawback.
Either they require a quantum computer for their implementation or they
have classical ciphertext.
In this work we aim for a QKR protocol that has all the desiderata one would expect:
\begin{itemize}[leftmargin=4mm,itemsep=0mm]
\item 
All actions on quantum states should be simple single-qubit actions like state preparation and measurement.
\item 
Alice should send only qubits, so that 
no bandwidth is wasted.
\item
Bob should send only an authenticated {\tt Accept}/{\tt Reject} bit. 
\item 
No key material should be consumed in case of {\tt Accept}, 
and the bare minimum\footnote{
For the noiseless case, the optimum is the length of the plaintext minus one \cite{DPS2005}.
}
should be consumed in case of {\tt Reject}.
\item 
The communication rate should equal that of QKD.
\end{itemize}

\section{Our Quantum Key Recycling protocol}
\label{sec:protocol}

\subsection{Protocol design considerations}

Our protocol is very similar to \cite{QKR_noise}. There are two main differences:

\begin{enumerate}[leftmargin=4.5mm,itemsep=0mm]
\item 
There is no classical communication from Alice to Bob.
\item 
In case of {\tt Reject} the keys are not thrown away. 
Instead, fresh key material is hashed into the old keys to obtain the keys for the next round.
\end{enumerate}

\begin{figure}[h]
\begin{center}
\includegraphics[width=.8\textwidth]{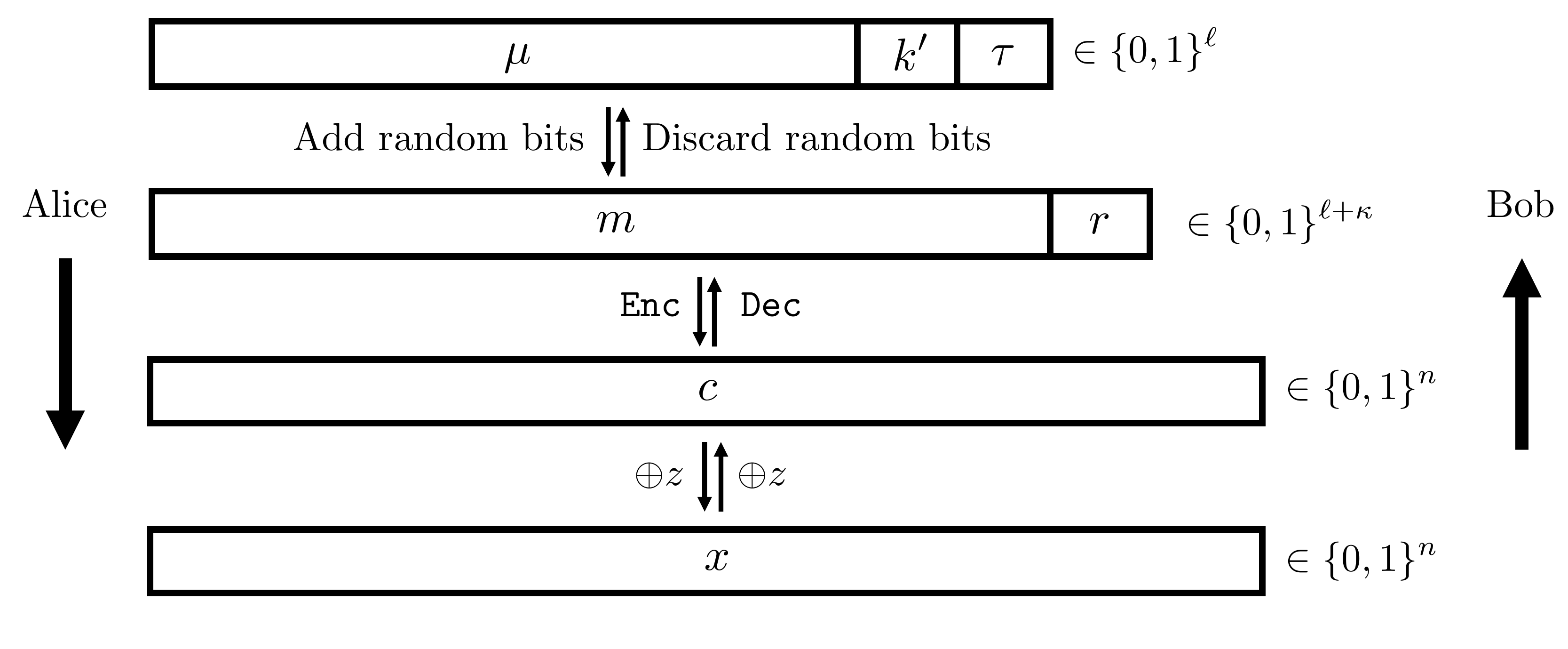}
\caption{\it 
Classical processing performed by Alice, and in reverse by Bob.
}
\label{fig:strings}
\end{center}
\end{figure}

In the transformation from \cite{QKR_noise} to a protocol without classical ciphertext,
there are several proof-technical issues.
Most importantly, the qubit payload $X\in\bits^n$ needs to be uniformly random.
(See Section~\ref{sec:permutations}. In the proof the $X$ acts as a uniform mask.)
This has to be reconciled with the fact that (i) the message is typically not uniform;
(ii) the error-correction encoding step introduces redundancy.
Our solution to these issues is shown in Fig.\,\ref{fig:strings}, which depicts most of the
variables in the protocol.
Alice first appends a random string $r \in \bits^\qk$ to the message, which will serve for privacy amplification.
Then she does the error correction encoding, resulting in a codeword~$c\in\bits^n$.
The $c$ is then masked with a one-time pad $z$; this masks any structure present in~$c$. 
A similar construction was proposed by Gottesman \cite{uncl}. 
However, instead of discarding $z$ we re-use most of the entropy in~$z$.

\subsection{Setup and protocol steps}

Alice and Bob have agreed on a linear error-correcting code with encoding and decoding functions 
${\tt Enc}:\bits^{\ell+\qk} \to\bits^n$ and 
${\tt Dec}: \bits^n \to \bits^{\ell+\qk}$.
The choice of $\qk$ and $n$ depends on the bit error rate of the quantum channel 
and on the required amount of privacy amplification.
Furthermore Alice and Bob have agreed on
a MAC function $\qG:\bits^\ql\times\bits^*\to\bits^\ql$, 
and two pairwise independent hash functions 
$F_u:\bits^n \times \cB^n\times \bits^\qk \to\bits^{n}\times\cB^n$ ($u\in\cU$) and
$G_v:\bits^n\times\cB^n\times\cQ\to\bits^{\ell}\times\cB^n$ ($v\in\cV$).\footnote{
The set $\cB$ is the alphabet of qubit basis choices.
In BB84 encoding we have $\cB=\{+,\times\}$; in 6-state encoding $\cB=\{x,y,z\}$.
}
The $\ql$ is a security parameter for the MAC function and is constant with respect to~$n$.

Alice's plaintext is $\mu\in\bits^{\ell-2\ql}$.
The key material shared between Alice and Bob consists of 
a mask $z \in \bits^n$, 
a MAC key $\xi \in \bits^\ql$ for Alice's message, 
a basis sequence $b \in \cB^n$, 
a MAC key $k\in \bits^\ql$ for Bob's feedback bit, 
and seeds $u\in\cU$, $v\in\cV$ for pairwise independent hashing.\footnote{
The strings $u$ and $v$ are never both used in the same round. 
We describe them independently since they have a different length, 
but the shorter ($v$) may as well be a substring of the longer ($u$).
} 
Furthermore Alice and Bob have a `reservoir' of additional spare key material.
\\

\begin{figure}[h]
\begin{center}
\includegraphics[width=\textwidth]{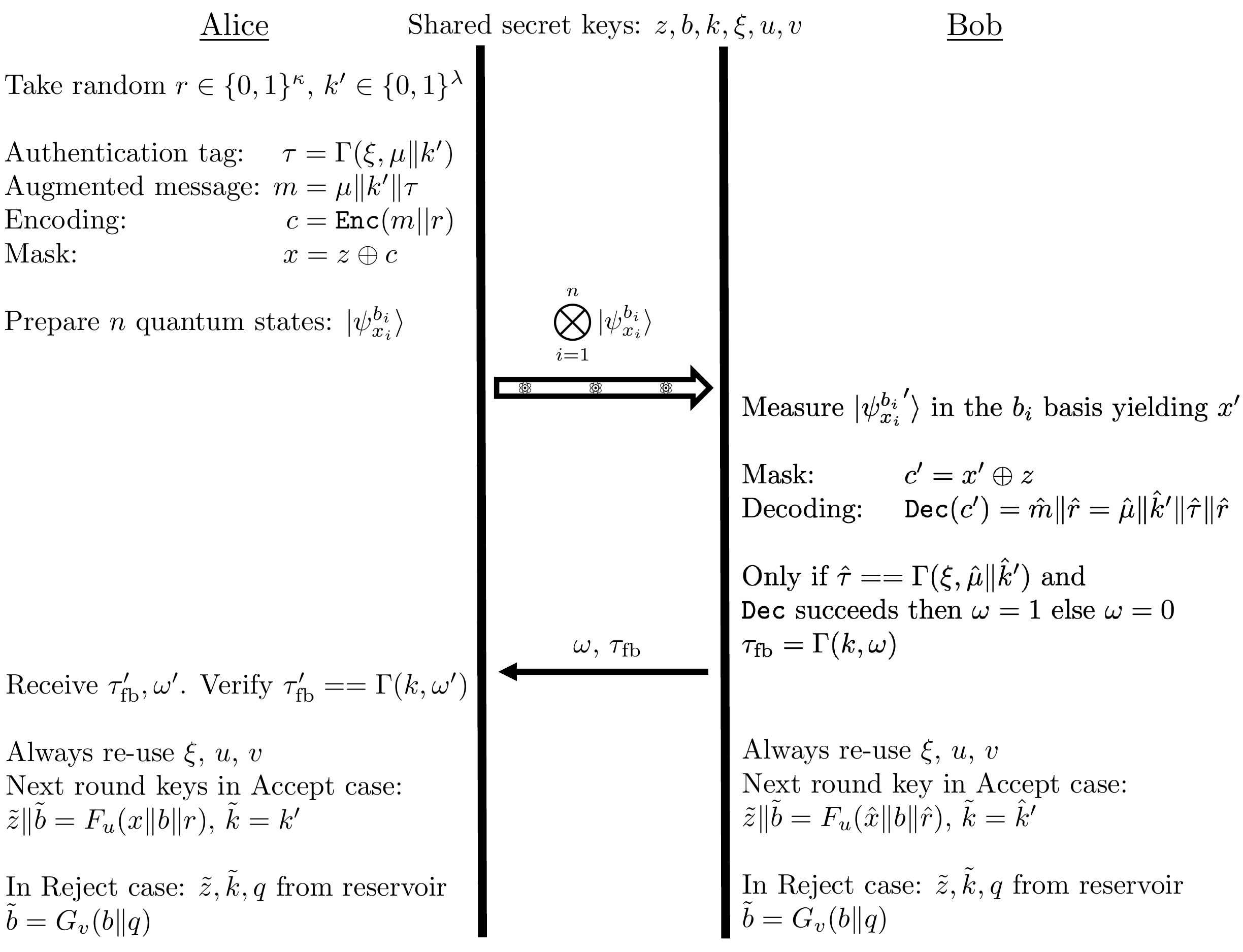}
\caption{\it 
One round of our 
QKR protocol. 
}
\label{fig:protocol}
\end{center}
\end{figure}

One round of the protocol consists of the following steps (see Fig.\,\ref{fig:protocol}):

\underline{Encryption}: \\
Alice generates random strings $r \in \bits^{\qk}$, $k' \in \bits^\ql$. 
She computes the authentication tag $\tau  = \qG(\xi, \mu\| k')$, 
the `augmented message' $m = \mu \| k' \|\tau$ (with $m\in\bits^\ell$),
the Encoding $c={\tt Enc}(m\|r)\in \bits^n$, 
and the qubit payload $x =c \oplus z$.
She prepares $\ket\qJ = \bigotimes_{i=1}^n\ket{\qj^{b_i}_{x_i}}$ and sends $\ket\qJ$ to Bob.\\
\underline{Decryption}:\\
Bob receives $\ket{\qJ'}$. He measures $\ket{\qJ'}$ in the basis $b$. The result is $x'\in\bits^n$. 
He computes $ c' =x' \oplus z$. 
He tries to recover $\hat m\| \hat r= {\tt Dec}(c')$.
If decoding does not generate a failure notification,
he discards $\hat r$ and parses $\hat m$ as $\hat m=\hat\mu \| \hat k' \| \hat\tau$.
He also computes $\hat c={\tt Enc}(\hat m\| \hat r)$
and $\hat x=\hat c\oplus z$.
\\
\underline{Feedback}:\\
Bob checks if $\qG(\xi,\hat \mu \|\hat k')== \hat\qt$. 
He sets $\qo = 1$ (`{\tt Accept}')
if the error correction did not fail and the MAC $\hat\qt$ is correct; 
$\qo=0$ (`{\tt Reject}') otherwise. 
He computes $\tau_{\rm fb} = \qG(k, \qo)$ and sends $\qo,\tau_{\rm fb}$ to Alice. 
Alice checks the MAC on the feedback.
\\
\underline{Key Update}:\\
The keys/seeds $\xi,u,v$ are always re-used.
The updated version of the $z,b,k$ in the next round is denoted as $\tilde z,\tilde b,\tilde k$.
\begin{itemize}[leftmargin=4mm,itemsep=0mm]
\item
In case of {\tt Accept}:\newline
Alice sets $\tilde k=k'$ and $\tilde z||\tilde b = F_u(x\|b\|r)$.\\
Bob sets $\tilde k=\hat k'$ and $\tilde z||\tilde b = F_u(\hat x \|b\|\hat r)$.
\item
In case of {\tt Reject}:\newline
Alice and Bob take new $\tilde z$ and $\tilde k$ from their reservoir.\\
They take $q\in\cQ$ from the reservoir and set
$\tilde b=G_v(b\| q)$.

\end{itemize}

\clearpage

\section{Protocol reformulation for the security proof}
\label{sec:equivalent}

We introduce a sequence of small modifications to the protocol of Section \ref{sec:protocol}.
While the original protocol $\cE_{\rm orig}$ in Section \ref{sec:protocol}
is the one that Alice and Bob actually execute,
we will write down the security proof for the modified protocol $\cE_{\rm mod}$. 
Due to their (almost-)equivalence,
security of $\cE_{\rm mod}$ implies security of $\cE_{\rm orig}$
up to a constant $2^{-\ql+1}$.
 
\begin{itemize}[leftmargin=4mm,itemsep=0mm]
\item
We mask the qubit payload with public randomness.
\item
We go to an EPR version in order to apply standard proof methods.
\item
We add random permutation of the qubits so that post-selection can be used.
\item
We add random Pauli transforms in order to simplify the noisy state.
\item
We pretend that the two authentication tags cannot be forged.
\end{itemize}

\subsection{Masking the qubit payload with public randomness}
\label{sec:publicmask}

Alice picks a random string $a \in \bits^n$. 
She computes $s=x\oplus a$.
Instead of qubit states $\ket{\qj^{b_i}_{x_i}}$ she prepares $\ket{\qj^{b_i}_{s_i}}$.
We denote Bob's measurement result as $t\in\bits^n$.
Alice publishes $a$ over an authenticated channel.\footnote{
This is a tamper-proof channel with perfect authentication.
}
Bob computes $x'=t\oplus a$.
Note that Eve learns $a$ only {\em after} she has attacked the qubits.
Since $a$ is public and independently random,
this roundabout way of getting $x'$ to Bob is equivalent to the original protocol
as far as security is concerned.

\subsection{EPR version of the protocol}
\label{sec:EPR}

Instead of having Alice prepare a qubit state and Bob measuring it,
now Eve prepares a noisy two-qubit EPR state (singlet state)
and gives the two subsystems `A' and `B' to Alice and Bob respectively.
Alice and Bob measure their $i$'th qubit in basis $b_i$;
this yields $s_i$ for Alice and $t_i$ for Bob, where 
$t_i$ equals $\overline{s_i}$ plus noise. 
The $s_i$ (or $t_i$) is random.

Alice computes $a=s\oplus x$ and publishes~$a$ in an authenticated way.
Bob computes $y=t\oplus a$. 
The rest of the classical processing is the same as in the original protocol,
with $x'=\bar y$.

Note that the statistics of the variables $s,t,a,x,x'$ is the same as in Section~\ref{sec:publicmask},
although the origin of the variables is now different.
The equivalence between prepare-and-measure on the one hand and
the EPR mechanism on the other hand has been
exploited in many works.

\subsection{Adding a random permutation}
\label{sec:permutations}

After Eve has handed out all $n$ EPR pairs, Alice and Bob publicly agree on a random permutation~$\pi$.
Before performing any measurement they both apply $\pi$ to their own set of $n$ qubits.
Then they forget~$\pi$.
The remainder of the protocol is as in Section \ref{sec:EPR}. 

For Alice and Bob the effect of the permutation is that the noise is distributed differently over the qubits.
The error-correction step is insensitive to the location of bit errors;
only the number of bit errors matters.
Hence all the classical variables that are processed/computed after the error correction step
are unaffected by~$\pi$. 
The only {\em output} variable of the protocol that is affected is~$a$.
However, $a$ was a uniform\footnote{
From Eve's point of view, $a$ gets randomized by $x$, which is uniform
because it is built from $z$, which is unknown to Eve. 
} 
random variable and has now become a different uniform variable;
as far as security is concerned,
the new protocol is equivalent to the one in Section~\ref{sec:EPR}.

Let $\cE_{\rm perm}$ denote the protocol containing the random permutation step.
In the language of Section~\ref{sec:post-selection} we can write
$\cE_{\rm perm}\circ\pi=\cE_{\rm perm}$. 
(After all, a permutation followed by a random permutation is a random permutation.)
We conclude that the post-selection criterion holds and we can apply~(\ref{eq:post-selection}).

Note that $\cE_{\rm perm}$ needs quantum memory.
This has no practical significance, since Alice and Bob actually execute $\cE_{\rm orig}$,
while $\cE_{\rm perm}$ is a proof-technical fiction.

\subsection{Adding random Pauli transforms}
\label{sec:Pauli}

This is the trick introduced by \cite{RennerThesis}.
For each individual EPR pair, Alice and Bob publicly agree on a random $\qa\in\{0,1,2,3\}$.
They both apply the Pauli transform $\qs_\qa$ to their own qubit state, and then forget~$\qa$.
This happens before they do their measurement.
The rest of the protocol is as in section \ref{sec:permutations}. 
The mapping in a single qubit position can be written as
\be
	\qr^{\rm AB}\mapsto  \tilde \qr^{\rm AB}=\fr14\sum_\qa (\qs_\qa\otimes\qs_\qa)\qr^{\rm AB} (\qs_\qa\otimes\qs_\qa).
\ee
The net effect of the Pauli transforms is that the measurement sequence $b$ gets 
randomized\footnote{
Let Alice and Bob both perform a projective measurement on their own part of $\tilde\qr^{\rm AB}$
in basis $\ket{\qj^b_x}, \ket{\qj^b_{\overline x}}$.
This can be rewritten as projective measurements on $\qr^{\rm AB}$
in basis $\qs_\qa\ket{\qj^b_x}, \qs_\qa\ket{\qj^b_{\overline x}}$.
}
with public randomness; but $b$ was already random, so security-wise nothing has changed.

The random-Paulis trick yields a major simplification: 
For six-state encoding (and higher),
only one degree of freedom is left in
the description of Eve's state, namely the bit error probability.
This was an important ingredient of the security proof in \cite{QKR_noise}.

\subsection{Pretending that the authentication tags are unforgeable}
\label{sec:auth}

We pretend that Eve is unable to forge the authentication tags
$\qt$ and $\qt_{\rm fb}$,
which is true except with probability $\leq 2\cdot2^{-\ql}$.
This has two benefits:
(i)
We get rid of complicated case-by-case analyses that would
allow events where the error correction yields a wrong $\hat m, \hat r$ without 
warning, while $\hat\qt$ looks correct;
(ii) 
In the {\tt Accept} case
Bob's reconstructed variables $\hat m,\hat r$
automatically equal Alice's $m,r$, thus reducing the number of variables.

\subsection{Effect of the modifications}
\label{sec:modeffect}

\begin{figure}[t]
\begin{center}
\includegraphics[width=1.0\textwidth]{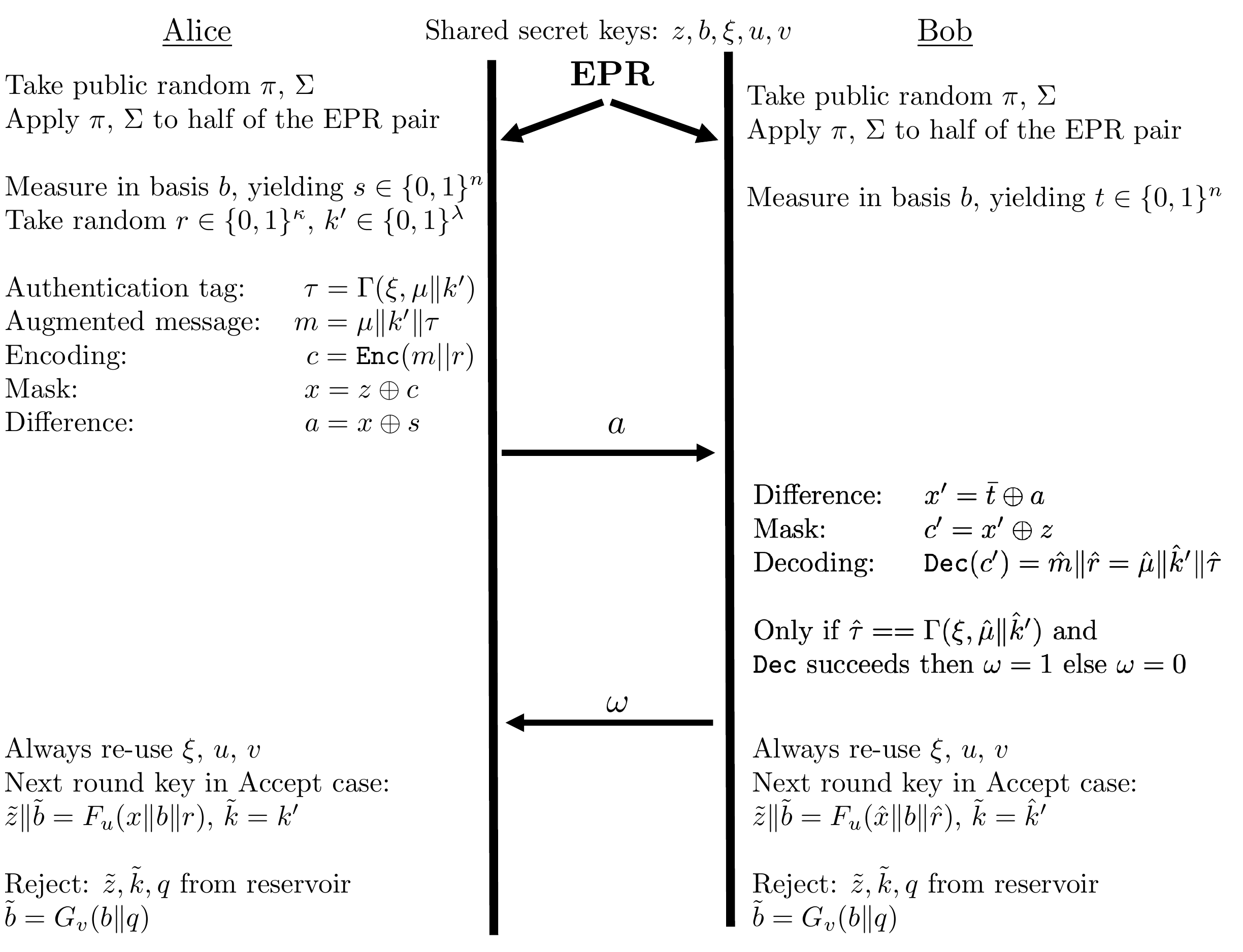}
\caption{\it 
The modified protocol with EPR states, random permutation, random Pauli transformations and perfect authentication. 
The notation $\pi$ stands for a permutation and $\Sigma$ for a vector of $n$ Pauli matrices.
}
\label{fig:equivalent}
\end{center}
\end{figure}

Fig.\,\ref{fig:equivalent} depicts the protocol $\cE_{\rm mod}$.
Due to the unforgeability of the tags we can write
\be
	\| \cE_{\rm orig}-\cF_{\rm orig} \|_\diamond \leq 2^{-\ql+1} + \| \cE_{\rm mod}-\cF_{\rm mod} \|_\diamond.
\ee
Furthermore, due to the permutation invariance of $\cE_{\rm mod}$ we can apply the post-selection proof technique
and use (\ref{eq:post-selection}).
Finally, thanks to the random Paulis,
the state $\qs$ in (\ref{eq:post-selection}) will have the very simple form
that makes it possible to arrive at an expression like
(\ref{OldAsymptotic}).

\section{The output state}
\label{sec:CPTP}

The Completely Positive Trace Preserving (CPTP) map $\cE_{\rm mod}$
acts on the `AB' subsystem (the $2n$ qubits controlled by Alice and Bob)
without affecting the `E' subsystem.
We write 
\be
	\cE_{\rm mod}=\cT\circ\cP\circ\cM\circ\cI.
\ee
The map $\cI$ fetches the classical input variables,
$\cM$ is the measurement, $\cP$ is the classical processing, and
$\cT$ traces away all variables that are not outputs.
The input variables are $mzbkuv$.
We have $\cI(\qr^{\rm ABE})=\EE_{mzbkuv}\ket{mzbkuv}\bra{mzbkuv}\otimes\qr^{\rm ABE}$.\footnote{
One can also start from a protocol description $\cE_{\rm mod}'$ that acts on a state
$\ket{\rm inputs}\bra{\rm inputs}\otimes \qr^{\rm AB}$, i.e. $\cE_{\rm mod}'$ describes how
the protocol acts on the quantum state $\qr^{\rm AB}$ given some value of the classical inputs.
The quantity of interest is then
$\cE_{\rm mod}'$ acting on a linear combination of input values;
this exactly matches the above mapping $\cI$.
}
Note that all input variables except $m$ are uniform.

The measurement $\cM$ introduces coupling between the classical $b$ register and the quantum state.
Furthermore, it destroys the AB subsystem and creates new classical registers $s,t\in\bits^n$.
\bea
	\cM \big(\ket b \bra b \otimes \qr^{\rm ABE}\big) = \EE_{st} \ket{bst}\bra{bst} \otimes \rho^{\rm E}_{bst}.
\eea
For the factorised form of $\qr^{\rm ABE}$ it holds that
$\EE_{st} (\cdots)=$
$\sum_{s t} 2^{-n} P_{t|s}(\cdots)$, with $P_{t|s} \isdef \qg^{|s\oplus \bar t|}(1-\qg)^{|s \oplus t|}$,
where $\qg$ is the bit error probability caused by Eve.

The processing $\cP$ introduces the new variables:\footnote{Here we do not keep track of the update $\tilde k$. Its security is trivial: it is updated either from $m$, which is confidential, or from the reservoir.} 
$r$ is generated randomly,
$cxay\qo \tilde z\tilde b$ are
created by Alice and Bob's computations and $q$ is fetched from the reservoir.
Let $n\qb$ be the number of bit errors that the error-correcting code can correct.
We define the indicator function $\qy_{st}$ such that $\qy_{st}=1$ when $|\bar s\oplus t|\leq n\qb$ and 
$\qy_{st}=0$ otherwise.
\bea
	(\cP\circ \cM \circ \cI)( \qr^{\rm ABE}) &=&
	\EE_{mzbkuvrq}\ket{mzbrkuvq}\bra{mzbkuvrq} 
	\otimes  \EE_{st} \ket{st}\bra{st} \otimes \rho^{\rm E}_{bst}
	\nn\\ &&
	\sum_{cxay\qo \tilde z\tilde b}
	\ket{cxay\qo \tilde z\tilde b}\bra{cxay\qo \tilde z\tilde b}
	\qd_{c,{\tt Enc}(m \| r)}  \qd_{x,c \oplus z} \qd_{a,x\oplus s}
	\nn\\ &&
	\qd_{y,t\oplus a}\qd_{\qo,\qy_{st}} \left[ \qy_{st}\qd_{\tilde z\|\tilde b, F_u(x\|b\|r)} 
	+\overline{\qy_{st}} 2^{-n}\qd_{\tilde b, G_v(b\|q)} 
	\right].
\label{stateAll}
\eea
The protocol output consists of the classical variables $a\qo m\tilde z\tilde b uv$.
The map $\cT$ traces out all the non-output registers.
Applying this trace to (\ref{stateAll}) yields
\bea
	\cE_{\rm mod}(\qr^{\rm ABE}) &=& 
	\qr^{UV\tilde Z\tilde B M A \qO \rm E}
	\nn\\ &=&
	\EE_{uvm\tilde z\tilde b a} \sum_{\qo} 
	\ket{uv\tilde z\tilde b ma\qo}\bra{uv\tilde z\tilde b ma\qo}
	\otimes
	[\qo  \qr^{{\rm E}[\qo=1]}_{u\tilde b\tilde z a}
	+\overline\qo  \qr^{{\rm E}[\qo=0]}_{v\tilde b a}]
	\quad\quad
\label{outputstate}
	\\
	\qr^{{\rm E}[\qo=1]}_{u\tilde b\tilde z a} &=&
	\EE_{bst} \rho^{\rm E}_{bst}\qy_{st} 
	2^n |\cB|^n \EE_r \qd_{\tilde z\|\tilde b,F_u[(s\oplus a) \|b\|r]}
\label{defrho1}
	\\
	\qr^{{\rm E}[\qo=0]}_{v\tilde b a} &=&
	\EE_{bst} \rho^{\rm E}_{bst}\overline{\qy_{st}} 
	|\cB|^n \EE_{q}\qd_{\tilde b,G_v(b \| q)}.
\label{defrho0}
\eea
In slight abuse of notation we have written 
$2^{-n}\sum_a=\EE_a$,
$2^{-n}\sum_{\tilde z}=\EE_{\tilde z}$,
$|\cB|^{-n}\sum_{\tilde b}=\EE_{\tilde b}$.
In (\ref{outputstate},\ref{defrho1},\ref{defrho0}) we should have formally written
$\qr^{{\rm E}[\qo=1]}_{uv\tilde b\tilde z ma}$ and $\qr^{{\rm E}[\qo=0]}_{uv\tilde b\tilde z ma}$,
but in the subscript we have kept only the variables on which the state actually has dependence.

The idealized version $\cF_{\rm mod}$ of the protocol is obtained by first executing $\cE_{\rm mod}$,
then tracing away the message $m$ and the keys $uv\tilde z\tilde b$, and finally replacing them
with completely random values.\footnote{
The distribution of $m$ does not have to be uniform.
}

\bea
	\cF_{\rm mod}(\qr^{\rm ABE}) &=& \chi^{UV\tilde Z\tilde B A}\otimes
	\EE_m \sum_\qo  \ket {m\qo}\bra{m\qo} \otimes
	\left(\qo \qr^{{\rm E}[\qo=1]} +\overline\qo \qr^{{\rm E}[\qo=0]}\right)
	\quad\quad
\label{Fmod}
	\\
	\qr^{{\rm E}[\qo=1]} &=& \EE_{bst}\qr^{\rm E}_{bst} \qy_{st}
	\\
	\qr^{{\rm E}[\qo=0]} &=& \EE_{bst}\qr^{\rm E}_{bst} \overline{\qy_{st}}.
\eea
The states with label `$[\qo=1]$' are sub-normalised; 
we have $\tr \qr^{{\rm E}[\qo=1]}=P_{\rm corr}$ and
$\EE_u\tr \qr^{{\rm E}[\qo=1]}_{u\tilde b\tilde z a}=P_{\rm corr}$,
where we define $P_{\rm corr}$ as
the probability that the number of errors can be corrected.
In the factorised form of $\qr$ it holds that
\be
	P_{\rm corr}(n,\qb,\qg)=\EE_{st}\qy_{st}
	= \sum_{c=0}^{\lfloor n\qb\rfloor} {n\choose c}\qg^c(1-\qg)^{n-c}.
\label{defPcorr}
\ee
Similarly $\tr \qr^{{\rm E}[\qo=0]}=1-P_{\rm corr}$ and
$\EE_v\tr \qr^{{\rm E}[\qo=0]}_{v\tilde b\tilde z a}=1-P_{\rm corr}$.

Note that when $m$ is uniform, the trace distance of the actual versus the ideal output state has an intuitive meaning
as the distance of the keys/seeds from uniformity given Eve's side information,
\bea
	\Big\| (\cE_{\rm mod}-\cF_{\rm mod})(\qr^{\rm ABE}) \Big\|_1 
	&=&
	\Big\| \qr^{UV\tilde Z\tilde B M A\qO \rm E} - \chi^{MUV\tilde Z\tilde B}\qr^{A\qO \rm E}\Big\|_1
	\\&=&
	2d(MUV\tilde Z\tilde B|A\qO \rm E). 
\eea

\section{Security Proof}
\label{sec:proof}

\subsection{Attacker Model}
The attacker model is the standard one in quantum cryptography. 
No information leaks from the labs of Alice or Bob, i.e.~there are no side-channels. 
Eve fully controls the environment outside Alice and Bob's labs.
Eve has unbounded quantum memory and unbounded (quantum-)computational resources.
Eve's measurements are noiseless.

\subsection{Forward secrecy}
\label{sec:resultforward}

Equations (\ref{outputstate}) and (\ref{Fmod}) serve as the starting point for the security proof.
Note that the expression (\ref{Fmod}) is also obtained if $M$ is {\em not} traced away;
consequently the analysis of known-plaintext and unknown-plaintext attacks turns out to be identical,
just as was the case in \cite{QKR_noise}.

An even stronger result holds: 
In (\ref{outputstate}) the $M$ is entirely decoupled from Eve's (classical and quantum) side information 
and from the next-round variables $UV\tilde Z\tilde B$.
Hence our protocol has forward secrecy:
a compromise of the updated keys has no impact on the secrecy of the message~$\mu$ \cite{Diffie1992}.

\subsection{Main result: upper bound on the diamond norm}
\label{sec:result}

\begin{theorem}
\label{th:main}
Let $\qr^{\rm ABE}$ have the factorised form $(\qs^{\rm ABE})^{\otimes n}$, with
$\qs^{\rm ABE}$ symmetrised by the random Pauli transform.
Let $\qe$ be a smoothing parameter, and let $\bar\qr$ denote a smoothed state.
Then
\bea
	&& \Big\|\cE_{\rm orig}-\cF_{\rm orig} \Big\|_\diamond <
	\nn\\ && \quad\quad 
	2^{-\ql+1}+
	(n+1)^{15}\bigg[
	 \frac1{2\sqrt{|Q|}}
	+\min\Big( P_{\rm corr}, \qe+\frac12{\rm tr}_{\rm E}\sqrt{ 2^{n-\qk} |\cB|^n{\rm tr}_{BS}(\bar\qr^{BSE})^2 } \Big)\bigg].
	\quad\quad
\eea
\end{theorem}

The $\min\{\cdots\}$ term is the same as in (\ref{DiamondIntermediate}) with $\ell$
replaced by $n-\qk$. 
This implies the asymptotic result (\ref{OldAsymptotic}) with $\ell$ replaced by $n-\qk$. 
Asymptotically it holds that the error correction redundancy has size $n-(\ell+\qk)\to nh(\qg)$.
This then yields an expression 
$\fr12\sqrt{2^{\ell-n+nh(\{1-\fr32\qg,\fr\qg2,\fr\qg2,\fr\qg2\})}}$.
We conclude that the asymptotic rate ($\ell/n$) equals 
$1-h(\{1-\fr32\qg,\fr\qg2,\fr\qg2,\fr\qg2\})$, 
as mentioned in Section~\ref{sec:prev_QKR}.

The term $\frac1{2\sqrt{|Q|}}$
dictates that, in order to have $\qa$ bits of security,
we have to set $\log|\cQ|>30\log(n+1)-2+2\qa$. 
Hence in case of {\tt Reject} the amount of expended key material is 
$n-1+30\log(n+1)+\ql+2\qa$. Asymptotically this is $n[1+\cO(\fr{\log n}n)]$.\\

\underline{\it Proof of Theorem~\ref{th:main}:}
The term $2^{-\ql+1}$ comes from the transition from $\cE_{\rm orig}$ to $\cE_{\rm mod}$.
The factor $(n+1)^{15}$ comes from applying the postselection theorem (\ref{eq:post-selection}).
For bounding the trace norm
$\|(\cE_{\rm mod}-\cF_{\rm mod})(\qr^{\rm ABE})\|_1$,
we start from (\ref{outputstate}),(\ref{Fmod}) and use the fact that
the eigenvalue problem reduces to an individual eigenvalue problem for each value of the classical variables,
orthogonal to the other values. 
We get
\bea
	\|(\cE_{\rm mod}-\cF_{\rm mod})(\qr^{\rm ABE})\|_1 &=& 
	D_{\tt acc}+D_{\tt rej}
	\\
	D_{\tt acc} &=& 
	\EE_{um\tilde z\tilde b a} \Big\|
	\qr^{{\rm E}[\qo=1]}_{u\tilde b\tilde z a}- \qr^{{\rm E}[\qo=1]} \Big\|_1
	\\
	D_{\tt rej} &=&
	\EE_{vm\tilde z\tilde b a} \Big\|
	\qr^{{\rm E}[\qo=0]}_{v\tilde b\tilde z a}- \qr^{{\rm E}[\qo=0]} \Big\|_1.
\eea
First we provide two upper bounds on $D_{\tt acc}$.
The first one simply follows from the triangle inequality,
\bea
	\EE_u\Big\| \qr^{{\rm E}[\qo=1]}_{u\tilde b\tilde z a}- \qr^{{\rm E}[\qo=1]} \Big\|_1
	&\leq&  \EE_u \| \qr^{{\rm E}[\qo=1]}_{u\tilde b\tilde z a} \|_1 
	+  \EE_u \| \qr^{{\rm E}[\qo=1]} \|_1
	\\&=& \EE_u \tr \qr^{{\rm E}[\qo=1]}_{u\tilde b\tilde z a}  +  \tr \qr^{{\rm E}[\qo=1]}
	=2P_{\rm corr}.
\eea
The second bound on $D_{\tt acc}$ takes some more work.
We introduce smoothing of $\qr$ as in \cite{RennerThesis,RK2005,TSSR2011}, allowing
states $\bar\qr$ that are $\qe$-close to $\qr$ in the sense of trace distance.
We have $D_{\tt acc}\leq 2\qe+\overline D_{\tt acc}$,
with $\overline D_{\tt acc}=$
$\EE_{uvm\tilde z\tilde b a} \|
\bar\qr^{{\rm E}[\qo=1]}_{u\tilde b\tilde z a}- \bar\qr^{{\rm E}[\qo=1]} \|_1$.
We write
\bea
	\overline D_{\tt acc} &=& 
	\EE_{mu\tilde z\tilde b a} \tr \sqrt{( \bar\qr^{{\rm E}[\qo=1]}_{u\tilde b\tilde z a}- \bar\qr^{{\rm E}[\qo=1]} )^2}
	\\ &\stackrel{\rm Jensen}{\leq}& 
	\EE_{m\tilde z\tilde b a}
	\tr \sqrt{\EE_u( \bar\qr^{{\rm E}[\qo=1]}_{u\tilde b\tilde z a}- \bar\qr^{{\rm E}[\qo=1]} )^2}
\label{eq:Jensen}
	\\ &=&
	\EE_{m\tilde z\tilde b a}
	\tr \sqrt{\EE_u( \bar\qr^{{\rm E}[\qo=1]}_{u\tilde b\tilde z a})^2 - (\bar\qr^{{\rm E}[\qo=1]} )^2}.
\label{Euvsquare}
\eea
In (\ref{eq:Jensen}) we used Jensen's inequality for concave operators.
In (\ref{Euvsquare}) we used 
$\EE_u \bar\qr^{{\rm E}[\qo=1]}_{u\tilde b\tilde z a} = \bar\qr^{{\rm E}[\qo=1]}$.
Next we evaluate the expression under the square root, making use of the 
properties of the pairwise independent hash function~$F$.
Squaring (\ref{defrho1}) yields
\bea
	&& \hskip-8mm
	\EE_u( \bar\qr^{{\rm E}[\qo=1]}_{u\tilde b\tilde z a})^2 - (\bar\qr^{{\rm E}[\qo=1]} )^2 
	\nn\\ && = 
	\EE_{bb'ss'tt'}\bar\qr^{\rm E}_{bst}\bar\qr^{\rm E}_{b's't'}\qy_{st}\qy_{s't'}
	2^{2n}|\cB|^{2n} \EE_{urr'} \qd_{\tilde z\|\tilde b,F_u[(s \oplus a) \|b\|r]} 
	\qd_{\tilde z\|\tilde b,F_u[(s' \oplus a) \|b'\|r']}
	- (\bar\qr^{{\rm E}[\qo=1]} )^2 
	\quad\quad  \\ && =  
	\EE_{bb'ss'tt'}\bar\qr^{\rm E}_{bst}\bar\qr^{\rm E}_{b's't'}\qy_{st}\qy_{s't'}
	\EE_{rr'}\! [ 1\!+\!(2^n |\cB|^n \!-\!1)\qd_{ss'}\qd_{bb'}\qd_{rr'} ] \!-\! (\bar\qr^{{\rm E}[\qo=1]} )^2
	\quad
	\\ && = 
	(2^n |\cB|^n -1) 2^{-\qk}\EE_{bb'ss'tt'}\qd_{bb'}\qd_{ss'}\bar\qr^{\rm E}_{bst}\bar\qr^{\rm E}_{b's't'}\qy_{st}\qy_{s't'}
	\\ && <  
	2^{n-\qk} |\cB|^n \EE_{bb'ss'tt'}\qd_{bb'}\qd_{ss'}\bar\qr^{\rm E}_{bst}\bar\qr^{\rm E}_{b's't'}
\label{thetalessone}
	\\ && =  
	2^{n-\qk} |\cB|^n \EE_{bb'ss'}\qd_{bb'}\qd_{ss'}\bar\qr^{\rm E}_{bs}\bar\qr^{\rm E}_{b's'}
	= 2^{n-\qk} |\cB|^n{\rm tr}_{BS}(\bar\qr^{BSE})^2.
\eea
In (\ref{thetalessone}) we used $\qy_{st}\leq 1$.
We have obtained the bound 
$\overline D_{\tt acc} <$ $\sqrt{2^{n-\qk} |\cB|^n}\cdot$
 ${\rm tr}_{\rm E}\sqrt{ {\rm tr}_{BS}(\bar\qr^{BSE})^2 } $.
We derive a bound on $D_{\tt rej}$ using similar steps, but without the smoothing.
Squaring (\ref{defrho0}) and taking the expectation $\EE_v$ we get
\bea
	&& \hskip-4mm
	\EE_v( \qr^{{\rm E}[\qo=0]}_{v\tilde b\tilde z a})^2 - (\qr^{{\rm E}[\qo=0]} )^2 
	\nn\\ && = 
	\EE_{bb'ss'tt'}\qr^{\rm E}_{bst}\qr^{\rm E}_{b's't'}\overline{\qy_{st}}\, \overline{\qy_{s't'}} 
	\EE_{qq'}|\cB|^{2n}\EE_v \Big[ \qd_{\tilde b,G_v(b \| q)} 
	\qd_{\tilde b,G_v(b' \| q')} \Big]
	- (\qr^{{\rm E}[\qo=0]} )^2
	\\ && =  
	\EE_{bb'ss'tt'}\qr^{\rm E}_{bst}\qr^{\rm E}_{b's't'}\overline{\qy_{st}}\, \overline{\qy_{s't'}} 
	\EE_{qq'}\Big\{ 1+(|\cB|^n-1)\qd_{bb'}\qd_{qq'}  \Big\}
	- (\qr^{{\rm E}[\qo=0]} )^2
	\\ && = 
	\frac{|\cB|^n-1}{|\cQ|}
	\EE_{bb'ss'tt'}\qd_{bb'}\qr^{\rm E}_{bst}\qr^{\rm E}_{b's't'}\overline{\qy_{st}}\, \overline{\qy_{s't'}} 
	\\ && < 
	\frac{ |\cB|^n}{|\cQ|} \EE_{bb'ss'tt'}\qd_{bb'}\qr^{\rm E}_{bst}\qr^{\rm E}_{b's't'}
	= \frac{ |\cB|^n}{|\cQ|}  \EE_{bb'}\qd_{bb'}\qr^{\rm E}_b \qr^{\rm E}_{b'} 
	= 
	\frac{1}{|\cQ|} (\qr^{\rm E})^2.
\eea
In the last step we used the special property that $\qr^{\rm E}_b$ does not actually depend on $b$
and thus equals $\qr^{\rm E}$ \cite{QKR_noise}.
(This property holds for the factorised and Pauli-symmetrised form of $\qr^{\rm ABE}$.)
We have obtained a bound $D_{\tt rej}<1/\sqrt{|Q|}$.
\hfill$\square$\\

In the proof above the updated $\tilde k$ does not appear explicitly.
The security of $\tilde k$ is guaranteed because 
(i) in the {\tt Accept} case the update resides inside $m$, which is secure;
(ii) in the {\tt Reject} case the update is done from the reservoir.

Similarly, in the proof the MAC key $\xi$ does not appear explicitly.
The fact that $m$ is secure implies that the tag $\qt$ remains confidential 
($\qt$ is a part of $m$), and hence there is no leakage about the MAC key $\xi$ that was
used to create the tag.

\section{Discussion}
\label{sec:discussion}

We have shown that the protocol in \cite{QKR_noise} can be modified in a way that 
{\em eliminates all classical communication from Alice to Bob, without increasing the number of qubits}. 
Essentially we have moved the classical OTP of \cite{QKR_noise} to the next QKR round.
Furthermore the error correction and authentication are happening `inside' the quantum state.
The asymptotic communication rate is not affected and is equal to the rate of QKD with one-way postprocessing.
Our protocol has forward secrecy.

The size of the keys shared by Alice and Bob is $n+n\log|\cB|+\log|\cU|+2\ql$,
(namely $z\in\bits^n$, $b\in\cB^n$,
$u\in\cU$, $\xi\in\bits^\ql$, $k\in\bits^\ql$),
with $\log |\cU|=n+n\log|\cB|$.
The size of $\cU$ could be reduced to $\log|\cU|\approx \qk$
by using {\em almost-pairwise independent} hashes. 

Another way of reducing the size of the initial key material is reducing the size of $z$ by $\qk$. 
In the encoding step we can choose to write $c$ in systematic form. 
Then a part of $c$ literally equals $r$, which is already uniform.
Since there is no need to mask an already uniform string, the size of $z$ and therefore the initial key material is reduced. 
Furthermore, the length of the seeds $u,v$ is reduced by~$\qk$.
The rate is unaffected by this modification. 

It is possible to take the seed $u$ from public randomness that is drawn in every QKR round.
This would not affect the security, and it would reduce the amount of shared key material.
However, it would require either 
(a) 
a source of public randomness that is not known by Eve beforehand, e.g. a broadcast;
or (b)
communication of $u$ from Alice to Bob or the other way round.
The former involves nontrivial logistics, while the latter violates the aims of this paper.

In the {\tt Accept} case the reservoir of shared key material remains untouched.
In the {\tt Reject} case the number of bits expended from the reservoir is
$n+\cO(\log n)$. 
Asymptotically, in the noiseless case ($n\to \ell+\qk$, $\qk\to 0$), 
this expenditure is very close to the optimum value $\ell$ \cite{DPS2005}.
(It is not possible to protect an $\ell$-bit message information-theoretically
with less than $\ell$ bits of key expenditure.)

We have not done anything about the classical feedback from Bob to Alice. 
It cannot be removed, because Alice needs to know if Bob correctly received her message.
On the other hand, one can consider a scenario where Alice and Bob are both senders,
in an alternating way.
Then the feedback bit can be placed inside the next message, resulting in a fully
quantum conversation.

\vskip2mm

There is one drawback to the protocol described in this paper. 
It is bad at dealing with erasures. 
As the actual message (as opposed to a random string) is encoded in the quantum state,
absorption of qubits in the quantum channel
has to be compensated in the error-correcting code.
The effect of erasures on the rate is severe. 
A solution as proposed in \cite{QKR_noise} would imply that the message is no longer encoded
directly in the qubits; instead Alice sends a random string to Bob, part of which survives the channel
and gets used to derive an OTP. 
Such a solution does not satisfy the aims of this paper.

As a topic for future work we mention finite-size analysis, e.g. smoothing without taking the
limit $n\to\infty$.

\subsection*{Acknowledgements}
Part of this research was funded by NWO (CHIST-ERA project ID\underbar{\;\;}IOT). 

\bibliographystyle{unsrt}

\bibliography{SilentBob}

\begin{thebibliography}{10}

\bibitem{BBB82}
C.H.\;Bennett, G.\;Brassard, and S.\;Breidbart.
\newblock {Quantum Cryptography II: How to re-use a one-time pad safely even if
  P=NP}.
\newblock {\em Natural Computing}, 13:453--458, 2014.
\newblock Original manuscript 1982.

\bibitem{DPS2005}
I.B.\;Damg{\aa}rd, T.B.\;Pedersen, and L.\;Salvail.
\newblock A quantum cipher with near optimal key-recycling.
\newblock In {\em CRYPTO}, pages 494--510, 2005.

\bibitem{DBPS2014}
I.B. Damg{\aa}rd, T.B. Pedersen, and L.~Salvail.
\newblock {How to re-use a one-time pad safely and almost optimally even if P =
  NP}.
\newblock {\em Natural Computing}, 13(4):469--486, 2014.

\bibitem{FehrSalvail2017}
S.~Fehr and L.~Salvail.
\newblock Quantum authentication and encryption with key recycling.
\newblock In {\em Eurocrypt}, pages 311--338, 2017.

\bibitem{QKR_noise}
D.~Leermakers and B.~\v{S}kori\'{c}.
\newblock {Security proof for Quantum Key Recycling with noise}.
\newblock Quantum Information \& Computation, 2019.
\newblock \url{https://eprint.iacr.org/2018/264}.

\bibitem{Barnum2002}
H.~{Barnum}, C.~{Crepeau}, D.~{Gottesman}, A.~{Smith}, and A.~{Tapp}.
\newblock Authentication of quantum messages.
\newblock In {\em The 43rd Annual IEEE Symposium on Foundations of Computer
  Science, 2002. Proceedings.}, pages 449--458, Nov 2002.

\bibitem{Portmann2017}
Christopher Portmann.
\newblock Quantum authentication with key recycling.
\newblock In Jean-S{\'e}bastien Coron and Jesper~Buus Nielsen, editors, {\em
  Advances in Cryptology -- EUROCRYPT 2017}, pages 339--368, Cham, 2017.
  Springer International Publishing.

\bibitem{uncl}
D.\;Gottesman.
\newblock Uncloneable encryption.
\newblock {\em Quantum Information and Computation}, 3(6):581--602, 2003.

\bibitem{SdV2017}
B.~\v{S}kori\'{c} and M.~de~Vries.
\newblock {Quantum Key Recycling with eight-state encoding. (The Quantum One
  Time Pad is more interesting than we thought)}.
\newblock {\em International Journal of Quantum Information}, 2017.

\bibitem{LS2018}
D.~Leermakers and B.~\v{S}kori\'{c}.
\newblock {Optimal attacks on qubit-based Quantum Key Recycling}.
\newblock {\em Quantum Information Processing}, 2018.

\bibitem{RK2005}
R.~Renner and R.~K\"{o}nig.
\newblock Universally composable privacy amplification against quantum
  adversaries.
\newblock In {\em Theory of Cryptography}, volume 3378 of {\em LNCS}, pages
  407--425, 2005.

\bibitem{TL2017}
M.~Tomamichel and A.~Leverrier.
\newblock A largely self-contained and complete security proof for quantum key
  distribution.
\newblock {\em Quantum}, 1:14, 07 2017.

\bibitem{WegmanCarter1981}
M.N. Wegman and J.W. Carter.
\newblock {New hash functions and their use in authentication and set
  equality}.
\newblock {\em Journal of computer and system sciences}, 22:265--279, 1981.

\bibitem{CKR2009}
M.~Christandl, R.~K\"onig, and R.~Renner.
\newblock Postselection technique for quantum channels with applications to
  quantum cryptography.
\newblock {\em Phys. Rev. Lett.}, 102:020504, Jan 2009.

\bibitem{RennerThesis}
R.~Renner.
\newblock {\em Security of quantum key distribution}.
\newblock PhD thesis, ETH Z\"{u}rich, 2005.

\bibitem{Diffie1992}
W.~Diffie, P.~Van~Oorschot, and M.~Wiener.
\newblock Authentication and authenticated key exchanges.
\newblock {\em Designs, Codes and Cryptography}, 2(2):107--125, Jun 1992.

\bibitem{TSSR2011}
M.~Tomamichel, C.~Schaffner, A.~Smith, and R.~Renner.
\newblock Leftover hashing against quantum side information.
\newblock {\em IEEE Transactions on Information Theory}, 57(8):5524--5535,
  2011.

\end{thebibliography}

\end{document}